\documentclass[
 aip,
 amsmath,amssymb,
reprint,%
nobalancelastpage
]{revtex4-2}

\usepackage{graphicx}
\usepackage{dcolumn}
\usepackage{bm}
\usepackage{siunitx}

\usepackage[utf8]{inputenc}
\usepackage[T1]{fontenc}
\usepackage{mathptmx}
\usepackage{etoolbox}
\usepackage[capitalize]{cleveref}

\makeatletter
\def\@email#1#2{%
 \endgroup
 \patchcmd{\titleblock@produce}
  {\frontmatter@RRAPformat}
  {\frontmatter@RRAPformat{\produce@RRAP{*#1\href{mailto:#2}{#2}}}\frontmatter@RRAPformat}
  {}{}
}%
\makeatother

\begin{document}

\preprint{AIP/123-QED}

\title{Millikelvin confocal microscope with free-space access and high-frequency electrical control}

\author{Thomas Descamps}
\altaffiliation[Now at: ]{Department of Applied Physics, KTH Royal Institute of Technology, Roslagstullsbacken 21, 10691 Stockholm, Sweden}
\affiliation{JARA-FIT Institute Quantum Information, Forschungszentrum Jülich GmbH and RWTH Aachen University, 52074 Aachen, Germany}
\author{Feng Liu}
\altaffiliation[Now at: ]{College of Information Science and Electronic Engineering, Zhejiang University, Hangzhou 310027, China}
\affiliation{JARA-FIT Institute Quantum Information, Forschungszentrum Jülich GmbH and RWTH Aachen University, 52074 Aachen, Germany}
\author{Tobias Hangleiter}
\affiliation{JARA-FIT Institute Quantum Information, Forschungszentrum Jülich GmbH and RWTH Aachen University, 52074 Aachen, Germany}
\author{Sebastian Kindel}
\affiliation{JARA-FIT Institute Quantum Information, Forschungszentrum Jülich GmbH and RWTH Aachen University, 52074 Aachen, Germany}
\author{Beata E. Kardyna\l}
\affiliation{Peter Grünberg Institute, Forschungszentrum Jülich GmbH, 52425 Jülich, Germany}
\affiliation{Department of Physics, RWTH Aachen University, 52074 Aachen, Germany}
\author{Hendrik Bluhm}
\affiliation{JARA-FIT Institute Quantum Information, Forschungszentrum Jülich GmbH and RWTH Aachen University, 52074 Aachen, Germany}
\email[]{bluhm@physik.rwth-aachen.de}

\date{\today}

\begin{abstract}
Cryogenic confocal microscopy is a powerful method for studying solid state quantum devices such as single photon sources and optically controlled qubits. 
While the vast majority of such studies have been conducted at temperatures of a few Kelvin, experiments involving fragile quantum effects often require lower operating temperatures. To also allow for electrical dynamic control, microwave connectivity is required. For polarization-sensitive studies, free space optical access is advantageous compared to fiber coupling. Here we present a confocal microscope in a dilution refrigerator providing all the above features at temperatures below 100 mK. The installed high frequency cabling meets the requirements for state of the art spin qubit experiments. As another unique advantage of our system, the sample fitting inside a large puck can be exchanged while keeping the cryostat cold with minimal realignment. Assessing the performance of the instrument, we demonstrate confocal imaging, sub-nanosecond modulation of the emission wavelength of a suitable sample and an electron temperature of 76 mK. While the instrument was constructed primarily with the development of optical interfaces to electrically controlled qubits in mind, it can be used for many experiments involving quantum transport, solid state quantum optics and microwave-optical transducers.
\end{abstract}

\maketitle

\section{Introduction}\label{sec:Introduction}

A confocal microscope operating at cryogenic temperatures is a key setup enabling the study of various quantum systems, such as molecular-beam-epitaxy grown quantum dots~\cite{Bayer2002,Greilich2011} and quantum wells~\cite{Schinner2013}, colloidal nanocrystals~\cite{Javaux2013}, molecules~\cite{Hwang2009}, 2D materials~\cite{Alexeev2019,Baek2020} and color centers~\cite{Robledo2011, Nguyen2019}. The in-depth understanding of these systems further allows the demonstration of a wide range of fundamental physics phenomena, including Rabi oscillation~\cite{Ramsay2010}, strong coupling~\cite{Yoshle2004,Qian2019}, photon blockage effect~\cite{Faraon2008} and Bose–Einstein condensation~\cite{Butov2002}, and new types of quantum devices/technologies, including on-demand quantum light sources~\cite{Ding2016,Liu2018}, few-photon all-optical switch/transistors~\cite{Sun2016,Englund2012}, optical quantum simulations~\cite{Wang2017} and quantum communications~\cite{Weber2019}.
However, most of the above experiments were performed at a few Kelvin. Lower temperature, e.g. $T<\qty{100}{\milli\kelvin}$, is highly desirable for further suppressing the noise caused by phonons or even a necessary condition for operating quantum bits (qubits) with energy level splittings $\lesssim\qty{100}{\micro\eV}$. In addition to the optical access, microwave connectivity is particularly useful for e.g. electrical dynamical control of a spin qubit~\cite{Joecker2019, Kuroyama2019} or detecting the spin state via the optically detected magnetic resonance technique~\cite{Clevenson2015}. Therefore, combining direct optical access with microwave connectivity and strong magnetic field at millikelvin temperatures not only improves the quality of the measurements and the performance of the quantum devices, but also open the possibility of many unprecedented experiments. A few examples are the implementation of a spin-photon interface~\cite{DeGreve2012, Gao2012, Bernien2013, Bhaskar2020} for gate-defined quantum dots with fully electrically controlled, long-lived spin qubits~\cite{Engel2006}, the development of microwave-optical quantum transducers \cite{Tsuchimoto2022, DeCrescent2022, Han2021, Weaver2023} to overcome the cryogenic limitations of a large number of superconducting qubits or the improvement of opto-mechanical sensors \cite{Li2021}. A confocal microscope constructed based on a dilution refrigerator equipped with high-frequency (RF) and DC cables and superconducting magnet could potentially fulfill all the above requirements.

To provide optical excitation without significantly increasing the base temperature, several architectures for dilution refrigerators have been reported. First with wet and then with dry dilution refrigerators, both free space-based \cite{Zrenner1998, Haupt2014, Kuroyama2019, Chervy2020} and fiber-based confocal microscopes have been designed.\cite{Hogele2008, Sladkov2011, Macdonald2015, Palmer2015, Evans2018} The latter provide better mechanical stability, but reduce the flexibility to control the polarization of the excitation laser and measure polarization/momentum-resolved photoluminescence (PL) spectra. Nowadays, dry dilution refrigerators including optical windows through the different shields for free-space propagation are available commercially. The optical windows can be incorporated on the sidewalls \cite{Morita2022, Onyszczak2023} (side access to the mixing chamber) or at the bottom of the lowest shields \cite{Chervy2020} (bottom access to the mixing chamber). The optics used for excitation, collection and imaging can be installed on an optical table next to/around the shields. However, the mechanical decoupling between the optics and the refrigerator makes the system more sensitive to vibrations. To remediate this problem, the second way consists of directly mounting a fiber-coupled optical cage system to the outer vacuum shield\cite{Chervy2020}. This cage system would probably have to be mounted/unmounted when opening/closing the fridge. Both of these approaches have in common that the optics cannot be pre-aligned before mounting all the shields. A third way to mount the optical head consists of placing it on top of the fridge \cite{Kuroyama2019, Evans2018}. It provides a better mechanical stability due to the direct coupling and the optics can be mounted permanently and aligned at room temperature when all the shields are dismounted.
Each of the systems mentioned above has its own capabilities to investigate and tune the sample properties: DC electrical lines, superconducting magnet, heaters and $N_{2}$ line to the mixing chamber for gas tuning. However, to the best of our knowledge, none of these systems offer high-frequency control. 
This is very limiting for (multi-)qubit experiments which need more channels, either DC or RF. One of the difficulties to incorporate high-frequency lines arises from the fact that it requires a rigid connection between semi-rigid coaxial cables to a movable sample mounted on piezo-stages at millikelvin temperatures. Besides, a significantly larger sample board to accommodate high-frequency components such as connectors, filters or bias-tees is desirable for optimal electrical signal routing and conditioning. 

In this work, we report the customization of a dry dilution refrigerator (Triton 450 from Oxford Instruments) with free-space optical access for confocal microscopy (section \ref{sec:confocal_microscope}), with high-frequency cables, a 2D vector magnet, heaters and filtered DC lines. Our system features a glass window on top of the cryostat and an optical head attached to the top plate of the fridge. A key element is the vertical placement of the sample board in conjunction with a static mirror to deflect the beam, along with high-load positioners sufficiently strong to move a board connected with several semi-rigid coaxial cables. The electrical wiring (section \ref{sec:Electrical_wiring}) allows state-of-the-art spin qubit measurements with microwave signals with 5 dB insertion loss at 5 GHz. Finally, the puck containing the objective and sample can be exchanged with the so-called “bottom-loading” technique, in our case provided by Oxford Instruments~\cite{Batey2014}. During this operation, the mixture circulation is stopped and the puck can be disconnected/reconnected to the mixing chamber (MC) plate while keeping the system under vacuum and cooled by the pulse tube refrigerator. Combined with the fixed optical head, this unique possibility of our system enables a fast sample turn-over with minimum optical realignment. We quantify the vibrations in our system by measuring the drift of the laser focused at the surface of a chip on the sample stage (section \ref{sec:vibrations}) and obtain a RMS-amplitude below 100 nm. Based on this setup, we demonstrate confocal imaging and dynamical electrical control of the emission wavelength of a gate-defined exciton trap with sub-nanosecond time scale (section \ref{sec:modulation}). Despite all the previous customization, the system can still efficiently cool as demonstrated by an electron temperature of \qty{76}{\milli\kelvin} in a gate-defined quantum dot (section \ref{sec:Electron temperature}).

\section{Confocal microscope\label{sec:confocal_microscope}}
\begin{figure}[tpb]
    \centering
    \includegraphics[width=\linewidth]{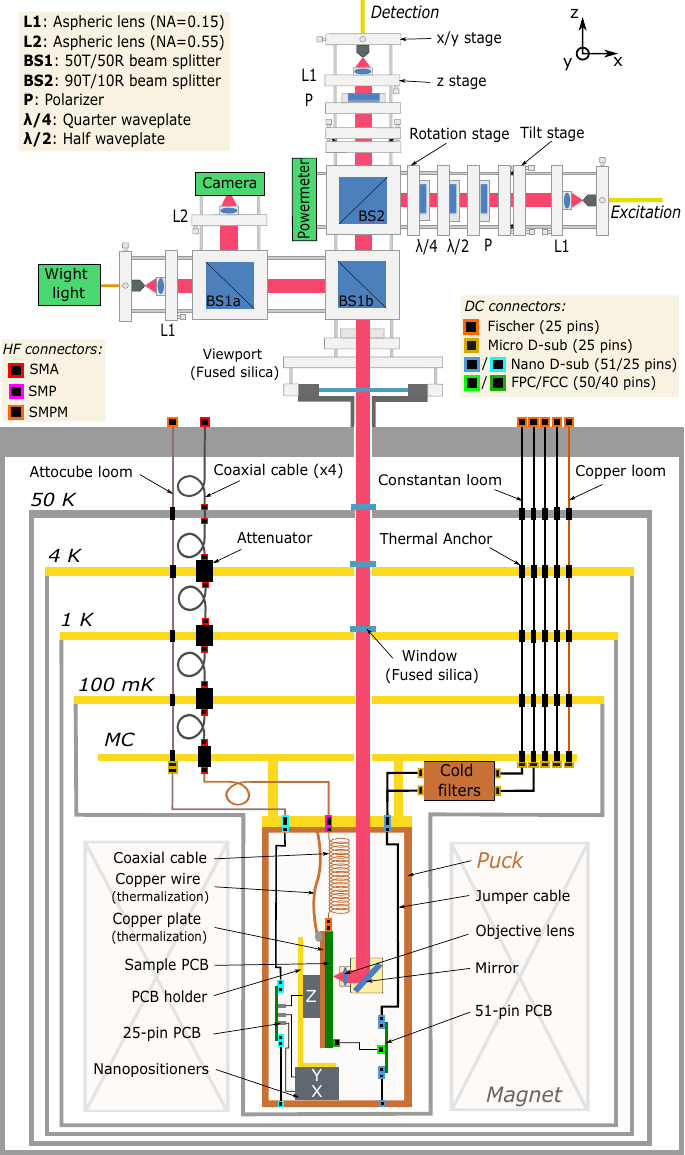}
    \caption{Schematic view of the confocal microscope. The parts are rescaled for clarity.}\label{fridge_sketch_setup}
\end{figure}

We now describe the confocal microscope that directs light towards the sample space and collects the returning emission. Direct reflection can be rejected from the emission signal by means of cross-polarization of the incident and detected laser beams.
A sketch of the full microscope can be seen in Fig.~\ref{fridge_sketch_setup}. All the parts inside the puck are non-ferromagnetic (made of copper, brass or titanium) to avoid displacement when a magnetic field is applied. The refrigerator rests on three air springs (CFM Schiller MAS 25) with resonance frequency of \qty{2.4}{\hertz} which are mounted on an Aluminium frame. The pulse tube refrigerator's (PTR's) rotary valve motor (not sketched) is mounted rigidly to the frame supporting the refrigerator in order to decouple the vibrations it generates from the fridge. Space constraints prohibit us from also mounting the cold head to a secondary reference frame and extending the flexible hose between it and the motor, which we expect would further reduce vibrations.\cite{Olivieri2017}

To transmit light into the cryostat, the top central blank port is replaced by a 15 mm diameter glass window in a DN40 CF flange. We chose fused-silica with an infrared coating for its almost perfect transmission in the near infrared and to attenuate external far-infrared radiation entering and thereby heating the cryostat. This flange is mounted to the top central port with a custom-made vacuum port.
We attached the optical head to the top of the cryostat in two ways to guarantee a rigid connection. First, the cage rods of the optical head are guided through an adapter (Thorlabs VFA275/M) tightened around the glass flange. Second, the cage rods are screwed to the custom-made vacuum port. The optical head does not have to be unmounted to open or close the cryostat. Besides, we observe negligible beam shifts during manipulation of the various shields of the cryostat, which confirms the stability of this configuration.
At each plate of the refrigerator, a hole (diameter 10 mm) was opened for free space light propagation inside the refrigerator. To avoid heat transfer between the plates by thermal radiation, fused-silica glass windows (Thorlabs WW41050-B) cover the holes of the \qty{50}{K}, \qty{4}{K} and \qty{1}{K} plates.

The optical head is designed for excitation, collection of the photoluminescence (PL) emission, and imaging of the sample mounted in the puck as shown on \cref{fridge_sketch_setup}.
The excitation laser light (near-infrared) coming from a single mode fiber (Thorlabs 780HP) is collimated by an aspheric lens (Thorlabs A280TM-B, $f = \qty{18.4}{mm}$, $\mathrm{NA} = 0.15$) on the right horizontal arm. A cube beam splitter BS2 (Thorlabs BS044) splits the beam so that \qty{10}{\percent} is reflected towards the cryostat and \qty{90}{\percent} is transmitted to the powermeter. 
Inside the puck, the excitation beam is reflected from a mirror at \qty{45}{\degree} and an aspheric objective lens (Thorlabs 355230-B, $f = \qty{4.51}{mm}$, $\mathrm{NA} = 0.55$) focuses it onto the sample mounted on a \qtyproduct{70 x 45}{mm} printed-circuit board (called "sample PCB"). As shown in Fig. \ref{fridge_sketch_setup}, the surface of the sample is parallel to the z-axis, which is also the direction of the beam propagating from room temperature to the puck. 
This sample orientation was chosen to maximize the in-plane magnetic field that can be supplied by the 2-axis vector magnet ($B_{z} = \qty{6}{T}$ and $B_{x} = \qty{3}{T}$). In addition, enough space is provided to mount the relatively large PCB. The diameter of the laser spot is on the order of \qty{3.5}{\micro m}. 
The reflected and PL signals from the sample are collected by the same objective lens and propagated along the same optical path as the excitation in the cryostat. At the optical head, the signals are transmitted through BS2 and finally collimated into a single mode fiber connected to the optical table for analysis. 
The confocal configuration improves the resolution by rejecting out-of-focus light. To image the sample, a white light source coming from a multimode fiber (0.10 NA, core diameter 105 µm) is collimated by an aspheric lens (Thorlabs C110TMD-B, $f = \qty{6.24}{mm}$, $\mathrm{NA} = 0.4$), first transmitted through a 50/50 beam splitter BS1a and then reflected towards the cryostat with a second 50/50 pellicle beam splitter BS1b (Thorlabs CM1-BP145B2). The reflected light from the sample is focused by an achromatic lens (Thorlabs AC254-100-B, $f = \qty{100}{mm}$) onto a CMOS camera (Thorlabs DCC1545M). If imaging of the surface is no longer needed, the pellicle beam splitter BS1b can be removed to improve the PL collection efficiency. To couple light into and out of the single mode fibers, each lens in the optical head fits inside a translation stage (Thorlabs SM1z) to adjust the focus while the fibers are mounted onto fine $x$/$y$ translation stages (Thorlabs CP1XY) to center the core along the optical axis. Finally, each arm is connected to a tilting stage (Thorlabs KC1-T/M) mounted to the cages holding the beam splitters in order to compensate angular displacement of the beams.
The sample PCB inside the puck is mounted on $xyz$ nanopositioner stages (Attocube ANPx311 with controller ANC350) to place it at the focal point of the objective lens and to move the sample at the desired excitation position. The nanopositioners are steppers with a travel range of \qty{6}{mm} and an accuracy of \qty{+- 200}{nm}. The three nanopositioners have ceramic bearings for additional rigidity against vibrations and the z-axis positioner holding the sample PCB is designed for high-loads. They can overcome the forces applied by the coaxial cables to the sample PCB which could potentially hamper the displacements.
The puck is firmly screwed to the MC plate for good thermal contact and can easily be taken in and out of the cold cryostat for sample exchange by the bottom-loading technique. After an exchange, the laser spot remains within the surface area of the sample and only fine adjustments of the beams with the optical head are sufficient to realign the system. We can implement the whole sample exchange procedure with realignment within a few hours, much faster than the few days that would take a full warm-up and cool-down.
A polarization system is mounted in the optical head to serve two purposes. 
One application is to excite the sample at a specific polarization and analyze the polarization of the emitted signals. For example, a linear polarizer (Thorlabs LPVIS050-MP2) mounted after the fiber mount of the excitation fiber linearly polarizes the excitation beam to an arbitrary angle. A half-wave plate (Bernhard Halle Nachfolger GmbH) mounted on a motorized rotation stage (Attocube ECR4040 with controller AMC100) can adjust the polarization to the desired angle. The sample can then be excited with linearly polarized light and the polarization of the emitted signal can be analyzed by mounting a polarizer on a motorized rotation stage in the detection arm before coupling to the fiber. If excitation with circular polarization is needed, a quarter-wave plate (Bernhard Halle Nachfolger GmbH) can be inserted in the optical path after the beam splitter BS1.
The second application is to reject the back-reflected laser light by cross-polarization \cite{Kuhlmann2013} for PL measurements under resonant excitation. A polarizer after the excitation fiber sets the polarization to an arbitrary angle and the following half-wave plate adjusts it to the desired value. In the detection arm, the polarizer is rotated to reject the polarization of the laser beam. A better extinction ratio can be obtained by adding a quarter-wave plate after the half-wave plate to compensate for the polarization distortion introduced by the optics. By iteratively rotating the half- and quarter-wave plate, rejection ratio around $10^{6}$ can be routinely achieved.

\section{\label{sec:Electrical_wiring} Electrical wiring}

\subsection{DC wiring}

\paragraph{Leads to the sample PCB}

The refrigerator was furnished with 4 Constantan looms (each made of 12 twisted pairs) intended to be used for DC measurements and thermally anchored at each plate to minimize heat load and thermal noise. Each loom is terminated by a 24-pin Fischer connector at room temperature and by a D-sub micro-D connector at the MC plate.

At room temperature, an RF-tight, home-built breakout box conditions the electrical signals applied to the sample before being transmitted to the looms. For each of the 96 lines, RC low pass filters (cut-off frequency \qty{17}{Hz} or \qty{1}{kHz}) to lower the noise, optional voltage dividers (1:6) to improve the voltage resolution or different shunts to ground (short, open, \qty{1}{\mega\ohm}) can be selected. Four heavily-shielded cables (from QDevil) link the break-out box to each of the four DC looms on top of the fridge.

\begin{figure}[tpb]
    \centering
    \includegraphics[width=\linewidth]{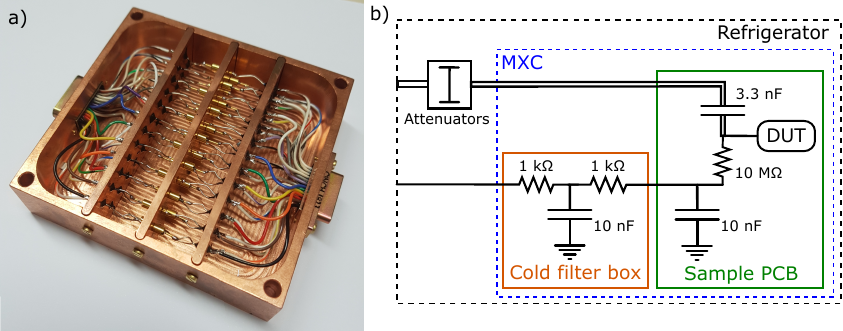}
    \caption{a) Cold filter box mounting on the MC plate filtering 24 DC lines. On the first and last panels, \qty{1}{\kilo\ohm} are glued and \qty{10}{nF} feedthrough resistors are screwed to the middle panel. b) Circuit schematic showing the DC filtering and the bias tee on the PCB sample. The single lines and the double line depicts the DC lines and the coaxial cable, respectively. }\label{DC_filtering}
\end{figure}

Inside the refrigerator, two looms are plugged to a homemade filter box made from OFHC copper screwed to the MC plate. A large contact area to the plate ensures proper thermalization and grounding. The box consists of two stacked compartments, each filtering 24 DC lines with 2nd-order RCRC filters (Fig. \ref{DC_filtering}). The connectors at the input and output of each compartment are D-sub micro-D connectors.
The \qty{1}{\kilo\ohm} resistors and \qty{10}{nF} feed-through capacitors are mounted to plates which have a low impedance electrical contact to the frame of the compartment. The resistors are glued into holes in a bulkhead mounting plate with epoxy (Stycast 2850FT) while the capacitors are screwed for proper thermalization and reduction of stray coupling bypassing the elements. Only the first three components of the 2nd-order filter are located in this box, and the last capacitor is soldered on the sample PCB in close proximity to the sample. A jumper cable consisting of 50 lines is connected to the output of the two compartments of the filter box while the other ending is terminated by a 51-pin nano-D screwed to the dock plate. 

Inside the puck, one jumper cable with nano-D connectors on both ends connect the mating 51-pin nano-D at the top of the puck to a PCB (called "51-pin PCB" in \cref{fridge_sketch_setup}). A second jumper cable connects the 51-pin nano-D at the bottom of the puck to the same PCB. On this 51-pin PCB, the lines of each jumper are routed to a 50-pin connector. A last flexible PCB connects the 50-pin connector to the 40-pin connector on the sample PCB mounted on the nanopositioners. This configuration conveniently avoids damage prone jumper cable manipulations when changing the sample. During the measurement, the top jumper is used to apply voltages to the sample while the lines on the bottom jumper are floating. During loading and unloading of the puck, the top jumper is not used while lines of the bottom jumper are grounded to protect the sample from electrostatic discharge. 

\paragraph{Leads to the nanopositioners}

One special loom for connecting the Attocube positioners was included inside the fridge with good thermal anchoring which features both low-resistance copper ($< \qty{5}{\ohm}$) and higher-resistance Constantan cables ($\approx\qty{200}{\ohm}$). The former are used to drive the nanopositioners in the puck and the latter to read out their positions. The connectors on both ends of this loom are the same as previously mentioned for the Constantan loom. A first jumper cable links the micro-D connector at the MC plate to a 25-pin nano-D connector screwed on the dock plate. A second jumper cable located inside the puck and terminated on both sides by a mating nano-D connector is plugged to a printed circuit board where fifteen of the lines are routed to three five-pin sockets (called "25-pin PCB" in \cref{fridge_sketch_setup}), each of them directly connected to one positioner. To reach the base temperature, the resistive readout of the positioners, which generates heat, is switched off. 

\subsection{AC wiring} \label{AC wiring fridge}
Semi-rigid coaxial cables, typically required for high-frequency control and readout of qubits, are now routinely mounted in dilution refrigerators.\cite{Batey2014, Rouxinol2016, Krinner2019} The material choice and cable diameter is a compromise between good electrical conductivity (to avoid signal loss) and poor thermal conductivity (to avoid thermal bridging between the plates). A comprehensive report of the choice of cables between each plate of the refrigerator can be found in Krinner \textit{et al.} [\onlinecite{Krinner2019}].
Inside the puck, thin copper semi-rigid cables (UT-34) connecting the sample PCB are shaped into spirals to further improve their flexibility and thus reduce the force applied on the nanopositioners. With around 20 loops of \qty{10}{mm} diameter, no additional transmission losses compared to a straight cable could be noticed. The cable has SMP female connector on the end connected to the dock plate of the puck and SMPM female on the other connected to the sample PCB. We implemented up to four cables with the current design, and we expect that potentially two or three more can be accommodated by having all three positioners designed for heavy load. 

To reduce the thermal noise introduced into the cables at room temperature (proportional to $k_{b}T$),\cite{Krinner2019} attenuators are included at each plate (except PT1, \textit{i.e.} the plate thermalized to the first stage of the pulse tube at $\approx 60$ K) with a good thermal contact. The heat generated by the signal dissipated in the attenuators is absorbed by the cooling power provided at each plate. To guarantee the proper operation of the refrigerator, the attenuation at each plate should therefore remain significantly below the maximum heat load sustainable by the plate. The total attenuation also depends on the signal amplitude needed to drive the samples. For spin-qubit operation, as the AC voltage amplitude required is on the order of a few tens of millivolt, the attenuation can be distributed as \qtylist[list-separator={/},list-final-separator={/},list-units=single]{20;6;6;3}{dB} among the plates (from PT2, \textit{i.e.} the plate thermalized to second stage the pulse tube at 4 K, to MC plate). However, for the purpose of one experiment described in Sec.~\ref{sec:modulation}, stronger AC signals were required and the attenuation was reduced to \qtylist[list-separator={/},list-final-separator={/},list-units=single]{6;0;0;3}{dB}. On the sample PCB, bias tees (made of a \qty{10}{\mega\ohm} resistance and a \qty{3.3}{nF} capacitor) were soldered to combine the high frequency and the DC signals, as depicted in \cref{DC_filtering}b). 

\section{Vibration performance\label{sec:vibrations}}
Ensuring low vibration at the sample stage is essential to carry out confocal microscopy, when the investigated emitter needs to be within the excitation spot which can reach one to a few microns. The PTR is an important source of periodic vibrations \cite{Schmoranzer2019} which have to be minimized at the sample to avoid an oscillatory misalignment of the excitation spot. In order to quantify the vibrations introduced by the pulse tubes at room temperature, we employed two different methods. First, we used a piezoelectric accelerometer (Wilcoxon 731-207) mounted at the sample position in the puck to measure the absolute displacement amplitude spectral density (ASD). Second, we measured the optical path's relative displacement ASD by a time-resolved reflection measurement of the laser focused at the sharp edge of a metal strip on a GaAs sample. While the former quantifies the absolute displacement ASD of a sample, the latter method allows us to determine the influence of vibrations on the optical performance of our system as it is sensitive to the relative displacements of the sample, the objective lens and all the individual parts of the optical head.

We mounted a GaAs sample whose surface was lithographically patterned with \qty{15}{\micro\meter} wide and \qty{200}{\micro\meter} thick Ti/Au metal gate in the puck. The laser was focused at the edge of the gate. The finite spot size of around \qty{3.5}{\micro\meter} effectively smooths the sharp transition between the gate and the GaAs surface so that the resulting reflectance gradient can be used as a sensitive probe of the relative lateral translation of the sample.
We then connected the detection arm of the microscope to an avalanche photodiode (Excelitas SPCM-AQRH) and digitized its output signal (Swabian Instruments Time Tagger 20). 
To relate the displacement perpendicular to the gate edge with the photon count rate, we 
performed a two-step calibration. First, we imaged the sample with the CMOS camera and white light and tracked the edge of the gate as a DC voltage of \qtyrange{0}{10}{\volt} is applied to the X and Y nanopositioners. By matching the resulting displacement in pixels on the camera to the known width of the gate, we obtained the magnification and thereby the displacement as a function of applied voltage. We then turned off the white light and measured the photon count rate as a function of voltage which yielded a linear slope over \qty{550}{\nano\meter}, resulting in a linear calibration of position versus count rate. 
We then positioned the sample in the middle of the slope and computed the displacement ASD from a time trace of the count rate using Welch's method.

\begin{figure}
    \centering
    \includegraphics{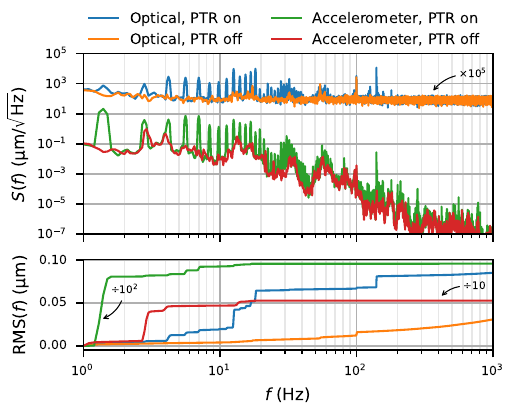}
    \caption{Vibration performance. The upper plot shows the vibration spectra measured using the two different methods described in the text. The lower plot shows the cumulative RMS up to frequency $f$. Some data are scaled by the indicated factors for clarity.}
    \label{fig:vibrations}
\end{figure}

\Cref{fig:vibrations} shows the ASD, $S(f)$, measured using the accelerometer and the reflectance method with the pulse tubes turned on and off. We reach a sensitivity (noise floor) of \qty{1}{\nano\meter\per\sqrt{\hertz}} with the optical method. Harmonics of the pulse frequency (\qty{1.4}{\hertz}) dominate the spectra for both methods. However, the frequency-resolved root-mean-square (RMS) amplitude with respect to the minimum frequency $f_\text{min}=\qty{1}{\hertz}$, 
\begin{equation}
    \text{RMS}(f) = \sqrt{\int_{f_\text{min}}^f\mathrm{d}{f'}\,S^2(f')}, 
\end{equation}
shown in the lower plot demonstrates that whereas the absolute displacement is dominated by the lowest-frequency peak, the optical path's susceptibility to vibrations is more uniform over frequency. This can be intuitively understood as the components along the optical path being excited in- phase by sources at low frequency, thus not contributing to relative displacement. Importantly to the optical capabilities of our setup, the overall RMS amplitude obtained from the optical measurement, $\text{RMS}(\qty{1}{\kilo\hertz})\approx\qty{100}{\nano\meter}$, is smaller by two orders of magnitude than the absolute level in the puck as measured with the accelerometer and well below the typical feature sizes of our samples. 
In contrast, displacement parallel to the beam direction was found to have negligible influence on the signal strength as we confirmed by applying a DC voltage to the focus axis nanopositioner.
We expect that more extensively decoupling the cold head and rotary valve motor \cite{Olivieri2017} as well as a stiffer puck and sample mount would further improve the performance.

\section{High-frequency modulation of the energy of a GaAs quantum well \label{sec:modulation}}

\begin{figure}[tpb]
    \centering
    \includegraphics[width=\linewidth]{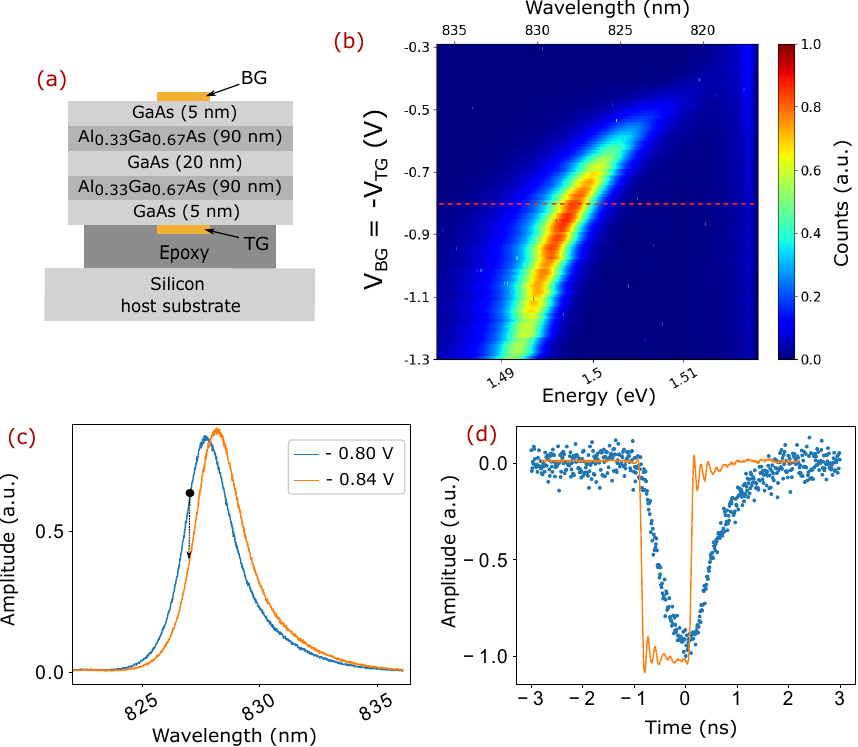}
    \caption{a) Sketch of the sample with local metal gates patterned on the two sides of the heterostructure. b) Stark shift of the quantum well photoluminescence as function of the bias voltage applied anti-symmetrically to the gates. c) Spectrum extracted at $V_{BG} = -V_{TG} = \qty{-0.8}{V}(\qty{-0.84}{V}$). During each voltage pulse (duration \qty{1}{ns} - effective amplitude \qty{80}{mV} - repetition rate \qty{150}{ns}) superimposed to the top-gate biased to \qty{0.8}{V}, a drop in the count rate is expected to occur if the gate voltage is originally set to \qty{0.8}{V}. This operating point is represented as the red dashed line in the color map. d) Optical pulse measured by the photon counting card (blue) and voltage pulse generated by the arbitrary waveform generator (orange). The two pulses have been normalized.}\label{pulsing}
\end{figure}

To test the high-frequency control in combination with the optical spectroscopy, we modulated the photoluminescence of an undoped GaAs quantum well with voltage pulses by the Stark effect. The sample layout is shown in \cref{pulsing}(a) and the fabrication is adapted from the process described by \citet{Descamps2023}.  

The sample was excited at \qty{790}{nm} by a continuous wave laser with \qty{100}{\micro\watt} power measured with the powermeter on the optical head (see \cref{fridge_sketch_setup}). The PL signal was dispersed by a spectrometer (Horiba FHR 1000) and imaged by a CCD camera (Andor iDus 416). By applying an anti-symmetric DC bias to the gates ($V_{BG} = -V_{TG}$), an electric field builds across the heterostructure, which shifts the PL emission of the quantum well to lower energy by the Stark effect as shown in \cref{pulsing}(b).
On top of the anti-symmetric DC bias $V_{BG} = -V_{TG} = \qty{-0.8}{V}$, an additional square pulse train is applied to the top gate. An arbitrary waveform generator AWG (Tektronix AWG70000B) was used to create the waveform: set peak amplitude \qty{125}{mV}, pulse width \qty{1}{ns} and repetition rate \qty{150}{ns}. By taking into account the attenuation of the line down to the PCB (\qty{10}{dB} at \qty{1}{GHz}) and the high impedance termination of the line, the pulse amplitude on the samples is approximately \qty{80}{mV} at the sample. 
Statically, applying \qty{80}{mV} only to the top-gate is equivalent to the application of \qty{40}{mV} to the top-gate and \qty{-40}{mV} to the back-gate in the configuration shown in \cref{pulsing}(b). Comparing the spectra at $V_{BG} = -V_{TG} = \qty{-0.8}{V}$ and $V_{BG} = -V_{TG} = \qty{-0.84}{V}$ (\cref{pulsing}(c)), we can see a small shift of the peaks but which results in a drop of the count rate at \qty{826.9}{nm}. To visualize the high-frequency shift of the quantum well emission, the dispersed PL spectrum was filtered at \qty{826.9}{nm} and the intensity was measured by an avalanche photodiode (Excelitas SPCM-AQRH). The time difference between the detected photons and the trigger signal from the AWG was recorded by a time tagger (Swabian Instruments Time Tagger 20) to obtain the correlation of these two signals. As shown in Fig. \ref{pulsing}(d), a dip in the normalized correlation is occurring at a voltage pulse. Compared to the shape of electrical pulse measured at the output of the source with \qty{7}{GHz} bandwidth oscilloscope, the optical pulses display a slower rise and fall times estimated to be \qty{0.5}{ns} by exponential fitting.
These slower rise and fall times originate mainly from the limited time resolution of the photo diodes ($\qty{0.35}{ns}$ from the datasheet) rather than the actual bandwidth of our high-frequency cabling. 

\section{Electron temperature \label{sec:Electron temperature}}

To assess potential heating due to stray radiation through the optical path of the microscope, we measured the electron temperature of a gate-defined quantum dot formed in the two dimensional electron gas of GaAs/AlGaAs \cite{Burkard2023, Hanson2007a}. To improve the thermal coupling of the sample to the MC plate, a plate with flexible copper wire connected to the top of the puck was inserted between the sample PCB and the PCB holder (see \cref{fridge_sketch_setup}).

\begin{figure}[tpb]
    \centering
    \includegraphics[width=0.9\linewidth]{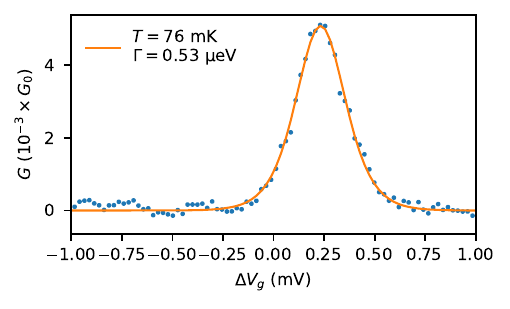}
    \caption{Electron temperature measurement. Conductance peak of a Coulomb resonance of a GaAs quantum dot in the sequential tunneling regime. The solid line is a fit to \cref{eq:coulomb_resonance} from which we obtain the electron temperature $T=\qty{76}{\milli\kelvin}$. $G_0 = 2e^2/h$ is the conductance quantum.}
    \label{fig:coulomb_resonance}
\end{figure}

We investigated the gate-defined quantum dot in transport measurements without laser excitation (but the optical windows were not covered). The DC voltages applied to the gates were supplied by a home-built, low-noise digital-to-analog converter (DAC) with \qty{16}{\bit} resolution. In addition to the cold filters, the signals were RC-filtered at room temperature with the home-built breakout-box (section \ref{sec:Electrical_wiring}) at a cut-off frequency of \qty{17}{\hertz}. The \qty{10}{\volt} output range of the DAC was further reduced by a factor of 6 using a resistive voltage divider. A bias voltage of $V_\mathrm{SD} = \qty{5}{\micro\volt}$ was applied to an Ohmic contact using the DAC and measured on the drain side using a Basel Precision Instruments SP 983C I/V converter. The quantum dot was tuned to the sequential tunneling regime where the temperature of the electron reservoir is the dominant energy scale and the measured Coulomb resonance was fitted to the function \cite{Ihn2010}
\begin{equation}\label{eq:coulomb_resonance}
    G(V_g) = \frac{e^2}{2h}\frac{\Gamma}{4k_\mathrm{B}T}\cosh^{-2}\left(\frac{\alpha (V_g - V_g^\mathrm{res})}{2k_\mathrm{B}T}\right).
\end{equation}
Here, $G(V_g)$ is the linear-response conductance through the dot as function of the gate voltage $V_g$ controlling the dot potential, $\Gamma$ is the tunneling rate between the dot and the reservoirs (assumed the same for source and drain). The coefficient $\alpha\approx \qty{0.08}{\milli\eV\per\milli\volt}$ is the leverarm, which converts the gate voltage to dot energy. It is obtained from a Coulomb diamond measurement as $\alpha = {E_C}/{\Delta V_g}$, with the charging energy $E_C$ and the voltage spacing between subsequent Coulomb resonances $\Delta V_g$ \cite{Hanson2007a}. We obtain a temperature of $T = \qty{76}{\milli\kelvin}$ and a tunneling rate of $\Gamma = \qty{0.53}{\micro\eV} \widehat{=} \qty{6}{\milli\kelvin} \ll T$, confirming we are indeed in the sequential tunneling regime (\cref{fig:coulomb_resonance}).

\section{Conclusion}

We presented the modification of a commercial dry dilution refrigerator allowing both free-space optical and high-frequency electrical access to a sample at millikelvin temperatures. Our configuration of a confocal microscope on top of the cryostat with a versatile puck enables fast sample turn-over with minimal optical realignment. We demonstrated electrical performance routinely achieved in solid-state quantum computing experiments exemplified by an electron temperature of \qty{76}{\milli\kelvin}, while at the same time being able to image and optically excite the sample as well as collect the PL signal. Possible applications of our system include the study of the magneto-optical properties of a variety of quantum emitters with advanced optical techniques, e.g. resonance fluorescence technique~\cite{Kuhlmann2013}, at a temperature an order of magnitude lower than the widely used \qtyrange[range-units = single,range-phrase = {--}]{1}{4}{\kelvin} optical cryostats. Our system can also be used to realize potential spin-photon interfaces and microwave-optical transducers operating at millikelvin temperatures.

\section{Acknowledgments}
We thank Marcus Esser for lending us the accelerometer.
We also thank Chao Zhao and Mihail Ion Lepsa for providing the heterostructure used in sections \ref{sec:vibrations} and \ref{sec:modulation} as well as Julian Ritzmann and Arne Ludwig for the one used in section \ref{sec:Electron temperature}. We thank Matthias K\"uenne for fabricating the chip used in section \ref{sec:Electron temperature}.

This work was funded by the European Research Council (ERC) under the European Union's Horizon 2020 research and innovation program (Grant agreement No. 679342), by the Deutsche Forschungsgemeinschaft (DFG, German Research Foundation) under Germany's Excellence Strategy - Cluster of Excellence Matter and Light for Quantum Computing (ML4Q) EXC 2004/1 -- 390534769, and by the Deutsche Forschungsgemeinschaft (DFG, German Research Foundation) -- 328514164.

\section{Author contributions}
The instrument was conceived by TD, FL, HB and BK. TD, FL, TH and SK built the microscope and characterized it. TD and TH analyzed the data. The project was supervised by HB and BK (supporting). TD wrote the original draft and all the authors contributed to reviewing and editing.

\bibliography{aipsamp.bbl}

\end{document}